\documentclass[
prl,
twocolumn,
superscriptaddress,
amsmath,
amssymb,
floatfix,
]{revtex4-2}

\usepackage[utf8]{inputenc}
\makeatletter
\let\latex@fnsymbol\@fnsymbol
\renewcommand\@fnsymbol[1]{\ensuremath{\ifcase#1\or * \or \dagger\or \ddagger\or
   \mathsection\or \mathparagraph\or \|\or **\or \dagger\dagger
   \or \ddagger\ddagger\or ***\or **** \else\@ctrerr\fi}}
\makeatother

\usepackage[usenames,dvipsnames]{xcolor}
\usepackage{booktabs,graphicx,mathrsfs,verbatim,soul}
\usepackage{textcomp}
\usepackage{float}
\usepackage{gensymb}
\usepackage{tabularx}
\usepackage{multirow}
\usepackage{dcolumn} 
\usepackage[mathlines]{lineno} 
\usepackage{bm} 
\usepackage{upgreek} 
\usepackage{multirow} 
\usepackage{makecell} 
\usepackage{placeins} 
\usepackage[colorlinks=true,citecolor=blue,filecolor=blue,linkcolor=blue,urlcolor=blue]{hyperref}
\usepackage{xspace}
\usepackage[caption=false]{subfig}
\usepackage[separate-uncertainty=true]{siunitx} 
\usepackage{appendix}
\usepackage{orcidlink}
\usepackage[export]{adjustbox}

\DeclareSIUnit\clight{\text{\ensuremath{c}}}
\DeclareSIUnit\evm{\eV\per\clight\squared}
\DeclareSIUnit\year{\text{y}}
\DeclareSIUnit\day{\text{d}}
\DeclareSIUnit\tonneyears{\tonne\times\year}
\DeclareSIUnit\tonnedays{\tonne\times\day}
\DeclareSIUnit\evm{\eV\per\clight\squared}
\DeclareSIUnit\gevm{\GeV\per\clight\squared}
\DeclareSIUnit\PE{\text{PE}}
\DeclareSIUnit\ph{\text{ph}}
\DeclareSIUnit\pperkev{\ph\per\kev}
\DeclareSIUnit\events{\text{events}}
\DeclareSIUnit\pertonneyears{\per(\tonne\times\year)}
\DeclareSIUnit\mev{\mega\eV}

\definecolor{summer}{RGB}{86,197,150}
\definecolor{pandablue}{HTML}{203769}
\definecolor{snogreen}{HTML}{105D20}

\newcommand{\cevns}{{\text{CE}\ensuremath{\upnu}\text{NS}}\xspace}

\newcommand{\tly}{\ensuremath{t_{\mathrm{Ly}}}\xspace}
\newcommand{\tqy}{\ensuremath{t_{\mathrm{Qy}}}\xspace}

\newcommand{\ly}{\ensuremath{\mathrm{L_\mathrm{y}}}\xspace}
\newcommand{\qy}{\ensuremath{\mathrm{Q}_\mathrm{y}}\xspace}
\newcommand{\lagr}{\mathcal{L}} 

\newcommand{\beight}{\ensuremath{^8\mathrm{B} }\xspace}
\newcommand{\ybe}{\ensuremath{^{88}\mathrm{YBe}}\xspace}
\newcommand{\krcal}{\ensuremath{^\mathrm{83m}\mathrm{Kr}}\xspace}
\newcommand{\kev}{\ensuremath{\mathrm{keV}}\xspace}

\newcommand{\fluxcmsec}{\ensuremath{\mathrm{cm}^{-2}\mathrm{s}^{-1}}\xspace}
\newcommand{\Sone}{$\mathrm{S1}$\xspace}
\newcommand{\Stwo}{\ensuremath{\mathrm{S2}}\xspace}

\newcommand{\arsq}{$^{37}$Ar\xspace}
\newcommand{\gevcsq}{\ensuremath{\mathrm{GeV}/c^2}}
\newcommand{\gevm}{\gevcsq}
\newcommand{\prevstwodt}{\ensuremath{\Stwo_\mathrm{pre}/\Delta t_\mathrm{pre}}\xspace}
\newcommand{\itsec}[1]{{\textit{#1} }---}

\newcommand{\nuis}{\theta}
\newcommand{\nuiss}{\vec{\nuis}}
\newcommand{\fref}[1]{Fig.~\ref{#1}}
\newcommand{\tref}[1]{Tab.~\ref{#1}}

\newcommand{\aref}[1]{Appendix~\ref{#1}}

\newcounter{appendix}
\renewcommand{\theappendix}{\Alph{appendix}}
\newcommand{\appendixsection}[1]{
    \refstepcounter{appendix}
    \textit{Appendix \theappendix: #1 ---} \label{app:appendix\theappendix}
}
\newcommand{\beightflux}{$(4.7_{-2.3}^{+3.6})\times 10^6$\,\fluxcmsec}

\begin{document}

\title{First Indication of Solar $^8$B Neutrinos via Coherent Elastic Neutrino-Nucleus Scattering with XENONnT}


\newcommand{\bologna}{\affiliation{Department of Physics and Astronomy, University of Bologna and INFN-Bologna, 40126 Bologna, Italy}}
\newcommand{\chicago}{\affiliation{Department of Physics, Enrico Fermi Institute \& Kavli Institute for Cosmological Physics, University of Chicago, Chicago, IL 60637, USA}}
\newcommand{\coimbra}{\affiliation{LIBPhys, Department of Physics, University of Coimbra, 3004-516 Coimbra, Portugal}}
\newcommand{\columbia}{\affiliation{Physics Department, Columbia University, New York, NY 10027, USA}}
\newcommand{\lngs}{\affiliation{INFN-Laboratori Nazionali del Gran Sasso and Gran Sasso Science Institute, 67100 L'Aquila, Italy}}
\newcommand{\mainz}{\affiliation{Institut f\"ur Physik \& Exzellenzcluster PRISMA$^{+}$, Johannes Gutenberg-Universit\"at Mainz, 55099 Mainz, Germany}}
\newcommand{\mpik}{\affiliation{Max-Planck-Institut f\"ur Kernphysik, 69117 Heidelberg, Germany}}
\newcommand{\munster}{\affiliation{Institut f\"ur Kernphysik, University of M\"unster, 48149 M\"unster, Germany}}
\newcommand{\nikhef}{\affiliation{Nikhef and the University of Amsterdam, Science Park, 1098XG Amsterdam, Netherlands}}
\newcommand{\nyuad}{\affiliation{New York University Abu Dhabi - Center for Astro, Particle and Planetary Physics, Abu Dhabi, United Arab Emirates}}
\newcommand{\purdue}{\affiliation{Department of Physics and Astronomy, Purdue University, West Lafayette, IN 47907, USA}}
\newcommand{\rice}{\affiliation{Department of Physics and Astronomy, Rice University, Houston, TX 77005, USA}}
\newcommand{\stockholm}{\affiliation{Oskar Klein Centre, Department of Physics, Stockholm University, AlbaNova, Stockholm SE-10691, Sweden}}
\newcommand{\subatech}{\affiliation{SUBATECH, IMT Atlantique, CNRS/IN2P3, Universit\'e de Nantes, Nantes 44307, France}}
\newcommand{\torino}{\affiliation{INAF-Astrophysical Observatory of Torino, Department of Physics, University  of  Torino and  INFN-Torino,  10125  Torino,  Italy}}
\newcommand{\ucsd}{\affiliation{Department of Physics, University of California San Diego, La Jolla, CA 92093, USA}}
\newcommand{\wis}{\affiliation{Department of Particle Physics and Astrophysics, Weizmann Institute of Science, Rehovot 7610001, Israel}}
\newcommand{\zurich}{\affiliation{Physik-Institut, University of Z\"urich, 8057  Z\"urich, Switzerland}}
\newcommand{\paris}{\affiliation{LPNHE, Sorbonne Universit\'{e}, CNRS/IN2P3, 75005 Paris, France}}
\newcommand{\freiburg}{\affiliation{Physikalisches Institut, Universit\"at Freiburg, 79104 Freiburg, Germany}}
\newcommand{\napels}{\affiliation{Department of Physics ``Ettore Pancini'', University of Napoli and INFN-Napoli, 80126 Napoli, Italy}}
\newcommand{\nagoya}{\affiliation{Kobayashi-Maskawa Institute for the Origin of Particles and the Universe, and Institute for Space-Earth Environmental Research, Nagoya University, Furo-cho, Chikusa-ku, Nagoya, Aichi 464-8602, Japan}}
\newcommand{\laquila}{\affiliation{Department of Physics and Chemistry, University of L'Aquila, 67100 L'Aquila, Italy}}
\newcommand{\tokyo}{\affiliation{Kamioka Observatory, Institute for Cosmic Ray Research, and Kavli Institute for the Physics and Mathematics of the Universe (WPI), University of Tokyo, Higashi-Mozumi, Kamioka, Hida, Gifu 506-1205, Japan}}
\newcommand{\kobe}{\affiliation{Department of Physics, Kobe University, Kobe, Hyogo 657-8501, Japan}}
\newcommand{\kit}{\affiliation{Institute for Astroparticle Physics, Karlsruhe Institute of Technology, 76021 Karlsruhe, Germany}}
\newcommand{\tsinghua}{\affiliation{Department of Physics \& Center for High Energy Physics, Tsinghua University, Beijing 100084, P.R. China}}
\newcommand{\ferrara}{\affiliation{INFN-Ferrara and Dip. di Fisica e Scienze della Terra, Universit\`a di Ferrara, 44122 Ferrara, Italy}}
\newcommand{\groningen}{\affiliation{Nikhef and the University of Groningen, Van Swinderen Institute, 9747AG Groningen, Netherlands}}
\newcommand{\westlake}{\affiliation{Department of Physics, School of Science, Westlake University, Hangzhou 310030, P.R. China}}
\newcommand{\shenzhen}{\affiliation{School of Science and Engineering, The Chinese University of Hong Kong, Shenzhen, Guangdong, 518172, P.R. China}}
\newcommand{\coimbrapoli}{\affiliation{Coimbra Polytechnic - ISEC, 3030-199 Coimbra, Portugal}}
\newcommand{\uniheidelberg}{\affiliation{Physikalisches Institut, Universit\"at Heidelberg, Heidelberg, Germany}}
\newcommand{\roma}{\affiliation{INFN-Roma Tre, 00146 Roma, Italy}}
\newcommand{\bucknell}{\affiliation{Department of Physics \& Astronomy, Bucknell University, Lewisburg, PA, USA}}
\author{E.~Aprile\,\orcidlink{0000-0001-6595-7098}}\columbia
\author{J.~Aalbers\,\orcidlink{0000-0003-0030-0030}}\groningen
\author{K.~Abe\,\orcidlink{0009-0000-9620-788X}}\tokyo
\author{S.~Ahmed Maouloud\,\orcidlink{0000-0002-0844-4576}}\paris
\author{L.~Althueser\,\orcidlink{0000-0002-5468-4298}}\munster
\author{B.~Andrieu\,\orcidlink{0009-0002-6485-4163}}\paris
\author{E.~Angelino\,\orcidlink{0000-0002-6695-4355}}\torino\lngs
\author{D.~Ant\'on~Martin\,\orcidlink{0000-0001-7725-5552}}\chicago
\author{F.~Arneodo\,\orcidlink{0000-0002-1061-0510}}\nyuad
\author{L.~Baudis\,\orcidlink{0000-0003-4710-1768}}\zurich
% ...
\collaboration{XENON Collaboration}\email[]{xenon@lngs.infn.it}\noaffiliation

\date{\today}

\begin{abstract}
We present the first measurement of nuclear recoils from solar $^8$B neutrinos via coherent elastic neutrino-nucleus scattering with the XENONnT dark matter experiment. The central detector of XENONnT is a low-background, two-phase time projection chamber with a 5.9\,t sensitive liquid xenon target. A blind analysis with an exposure of 3.51\,t$\times$yr resulted in 37 observed events above 0.5\,keV, with ($26.4^{+1.4}_{-1.3}$) events expected from backgrounds. The background-only hypothesis is rejected with a statistical significance of 2.73\,$\sigma$. The measured $^8$B solar neutrino flux of $(4.7_{-2.3}^{+3.6})\times 10^6\,\mathrm{cm}^{-2}\mathrm{s}^{-1}$ is consistent with results from the Sudbury Neutrino Observatory. The measured neutrino flux-weighted CE$\upnu$NS cross section on Xe of $(1.1^{+0.8}_{-0.5})\times10^{-39}\,\mathrm{cm}^2$ is consistent with the Standard Model prediction. This is the first direct measurement of nuclear recoils from solar neutrinos with a dark matter detector.
\end{abstract}

\maketitle

\itsec{Introduction}
Coherent elastic neutrino-nucleus scattering\,(\cevns) is a Standard Model\,(SM) process with low momentum transfer, which allows neutrinos to scatter coherently with nuclei~\cite{Freedman:1973yd,Kopeliovich:1974mv,Drukier:1984vhf}. This process has only recently been observed using an intense, pulsed 
spallation neutron source~(SNS)~\cite{COHERENT:2017ipa,COHERENT:2020iec}. The detection of \cevns events from solar neutrinos is more challenging due to the lower flux~\cite{Bahcall:2004mz} and energy, as well as the lack of timing information. Therefore, it requires minimal backgrounds and maximizing the sensitive region of interest\,(ROI) with a low-energy threshold. Liquid xenon\,(LXe) detectors searching for dark matter\,(DM) particles fulfill these requirements, but have not been able to reach the required sensitivity until now~\cite{xenon1t_b8,PandaX:2022aac-b8}. Solar $^8$B neutrinos are expected to contribute the largest detectable number of coherent neutrino-xenon scattering events, albeit at low nuclear recoil\,(NR) energies~\cite{Strigari:2009bq}. In this Letter, the first detection of \cevns induced by solar \beight neutrinos with the XENONnT experiment is reported. This is a ``first'' in three different aspects: the first detection of elastic NRs from astrophysical neutrinos, the first measurement of the \cevns process with a Xe target, and the first step into the ``neutrino fog'' by a DM experiment~\cite{Billard:2013qya,OHare:2021utq}. 

\itsec{Experiment}
The XENONnT experiment~\cite{XENON:2024wpa}, located at the INFN Laboratori Nazionali del Gran Sasso in Italy, is designed to search for weakly interacting massive particles\,(WIMPs) scattering off Xe nuclei, which has a similar NR signature as \cevns. The experiment consists of three nested detectors: a muon veto\,(MV), a neutron veto\,(NV), and an innermost LXe detector. The latter is a two-phase time projection chamber\,(TPC) housed in a double-walled cryostat filled with 8.5\,t of LXe.  The cylindrical TPC, 1.33\,m in diameter and 1.49\,m in height, is enclosed by polytetrafluoroethylene\,(PTFE) panels and viewed by 494 3-in. Hamamatsu R11410-21 photomultiplier tubes\,(PMTs)~\cite{Antochi:2021wik} arranged in a top and a bottom array. The active LXe mass in the TPC is 5.9\,t. 

Particle interactions in the TPC produce both scintillation photons and ionization electrons. The prompt scintillation photons are detected by the PMTs and are referred to as the S1 signal. The liberated electrons drift upward in the drift field to the liquid-gas interface, where they are extracted into the gas and produce a secondary scintillation signal, called the S2 signal, via electroluminescence. The time difference between \Sone and \Stwo signals is proportional to the interaction depth\,($Z$). Event positions in the horizontal plane\,($X, Y$) are reconstructed based on the hit patterns of S2 signals in the top PMT array. 

The electric fields in the TPC are established by three parallel-wire electrodes made of stainless steel~\cite{XENONnT:2023dvq}. The cathode and gate electrodes establish a drift field at 23\,V/cm, resulting in a maximum electron drift time of \SI{2.2}{\milli\second}. The extraction field in LXe is set to 2.9\,kV/cm by the gate and anode electrodes, which are reinforced by two and four additional perpendicular wires, respectively, to minimize sagging~\cite{XENONnT:2023dvq}. Two additional parallel-wire electrodes shield the PMT arrays from electric fields~\cite{XENON:2024wpa}.

\itsec{Dataset} This search uses two datasets with a total live time of 316.5\,days after accounting for dead time from data acquisition~\cite{XENON:2022vye} and vetoes. The first dataset, taken between July 6, 2021 and November 28, 2021 is referred to as the SR0 dataset in this Letter with a live time of 108.0\,days. The second dataset was collected between May 19, 2022 and August 8, 2023, a period referred to as SR1, with a live time of 208.5\,days. During SR0 (SR1), the temperature and pressure in the detector are stable within (176.8$\pm$0.4)~[(177.2$\pm$0.4)]\,K and (1.890$\pm$0.004)~[(1.92$\pm$0.02)]\,bar, respectively. The liquid level in SR0 is stable within (5.02$\pm$0.20)\,mm~\cite{XENON:2021qze}. On July 15, 2022, the liquid level is lowered by 0.2\,mm and the anode voltage is raised by 50\,V to mitigate localized electron bursts and maintain a consistent extraction field strength. Before and after this adjustment, the liquid level in SR1 is maintained stable at 5.0 and 4.8\,mm above the gate electrode, respectively.
The systematic uncertainty of the liquid level measurement is 0.2\,mm.

In addition to the 17 PMTs already excluded from the analyses during SR0~\cite{XENON:2024qgt}, three additional PMTs are removed in SR1 due to increased afterpulsing or intermittent light emission. PMT gains are monitored weekly using pulsed LED signals, and are found to be stable in SR0 (SR1) within 3\%~(3.5\%). PMT hits are recorded on a per-PMT basis when crossing the digitization threshold, typically about 2.06\,mV~\cite{XENON:2022vye}. The mean single photoelectron\,(PE) acceptance in SR0 (SR1) is determined to be (91.2$\pm$0.2\%)~[(92.1$\pm$0.7)\,\%]. Clusters of PMT hits in time are divided into peaks, which are classified into S1 and S2 signals based on their waveforms and intensity distributions on PMT arrays~\cite{strax,straxen}.

A distortion of the drift field near the edges of the detector leads to a difference in positions between the interaction site and the extraction position. It also leads to a small charge-insensitive volume (CIV)~\cite{XENONnT:2023dvq} in the lower part of the TPC, from where the drifting electrons reach the PTFE wall instead of the liquid-gas interface. A data-driven correction for the radial coordinate is applied to reproduce the uniform distribution of \krcal calibration events~\cite{XENON:2019ykp}. For SR0, the method from~\cite{XENON:2022ltv} is kept, where the CIV does not enter the position correction but is considered in the fiducial volume~\,(FV) calculation. The FV mass uncertainty originated from field distortion and position reconstruction is less than 5\%. In SR1, the event positions are corrected according to the boundary defined by the simulated drift field~\cite{XENONnT:2023dvq} to account for the CIV. After considering the field distortion correction and removing events with the interaction depth $Z$ below -142 cm or above -13 cm due to an insufficient understanding of the detector and backgrounds, the FV mass for SR0 and SR1 are (3.97 $\pm$ 0.20) and (4.10 $\pm$ 0.19)\,t, respectively. The total exposure in this analysis is 3.51\,t$\times$yr.

Light from S1 or S2 signals can create delayed electron signals via photoionization of impurities in the LXe~\cite{XENON:2021qze}. The photoionization strength, defined as the ratio between the number of measured photoionization electrons within \SI{2.2}{\milli\second} after an \Stwo signal larger than \SI{10000}{\PE} and the number of electrons in the \Stwo signal itself, increased tenfold after a long maintenance and upgrade phase between SR0 and SR1. One hypothesis of the increased photoionization is that components in the radon removal system \cite{Murra:2022mlr} are releasing photoionizable impurities after the upgrade, which enabled high flow extraction from the LXe target. No impact is observed from these impurities on the electron lifetime, which is an attenuation coefficient for the attachment to electronegative impurities during the drift of ionization electrons.

Signal inhomogeneities due to position- and time-dependent effects are corrected as described in \cite{XENON:2022ltv}. The increased and varying photoionization strength in SR1 requires further time-dependent corrections to the \Sone and \Stwo signal areas. 
After all corrections, the stability of the corrected \Sone and corrected \Stwo (c\Stwo) signals in SR0 are within 1\% and 1.9\%, respectively, and 0.3\% and 1.1\% in SR1. The variations are propagated as uncertainties into the determination of the photon gain~($g$1) and electron gain~($g$2). Using the method described in \cite{XENON:2022ltv}, $g$1 and $g$2 in SR0 (SR1) are found to be (0.151$\pm$0.001)~[(0.137$\pm$0.001)]\,PE/photon and (16.5$\pm$0.6)~[(16.9$\pm$0.5)]\,PE/electron, respectively.

\begin{figure}[t]
    \centering
    \includegraphics[width=0.95\columnwidth,left]{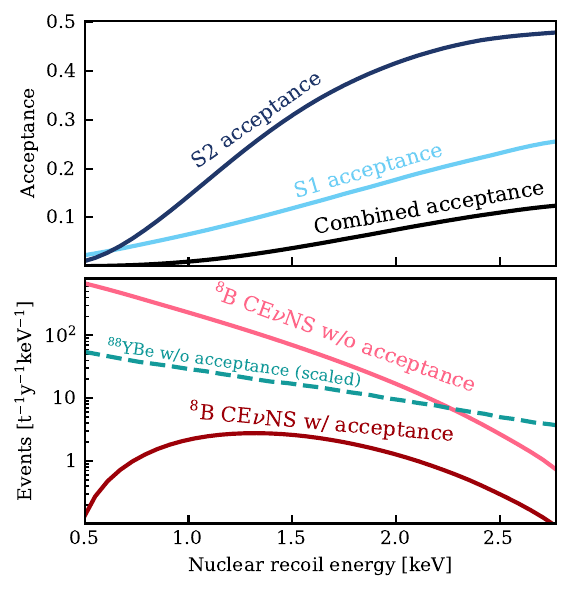}
    \caption{Acceptance for detecting low-energy NRs in XENONnT\,(top) and \beight~\cevns energy spectrum\,(bottom). The light (dark) blue curve denotes the acceptance of detecting S1 (S2) signals, and the black curve represents the combined acceptance. The expected \cevns signal spectrum induced by solar \beight neutrinos in XENONnT with (without) acceptance loss is shown by the dark (light) red line. The green line shows a scaled spectrum of all energy depositions from \ybe calibration.}
    \label{fig:detection_acceptance}
\end{figure}

\itsec{\cevns signal}
The expected NR spectrum of \beight~\cevns in LXe, considering the solar \beight neutrino flux measured by the Sudbury Neutrino Observatory(SNO)~\cite{SNO:2011hxd}, the \beight neutrino energy spectrum from~\cite{Bahcall:1996qv} and the \cevns cross section on Xe predicted by the SM~\cite{Barranco:2005yy}, is shown in Fig.~\ref{fig:detection_acceptance}, with 90\% of detectable recoils between 0.7 and 2.1\,keV. The main contribution is from neutrinos with energies between 8 and 15\,MeV. The low-energy NR response in this search is calibrated with 152\,keV neutrons from an external \ybe source~\cite{Collar:2013xva}, with the recoil spectrum also shown in\,\fref{fig:detection_acceptance}. The uncertainty in signal acceptance arises from uncertainties in S1 reconstruction, classification acceptance, and event selection acceptance. A model for light yield\,(\ly) and charge yield\,(\qy) is fitted~\cite{appletree} to calibration data using a method similar to that described in \cite{XENON:2024xgd}. The uncertainties of yields are propagated into the final inference with two parameters, $\tly$ and $\tqy$, which determine the relative shift of \ly and \qy from their median toward the $\pm$1$\sigma$ quantiles. This calibration will be presented in an upcoming publication~\cite{YBE_inprep}. \ly and \qy below~\SI{0.5}{\kev} are assumed to be zero, which has a negligible impact on the \beight~\cevns~detection rate.

The expected \beight\,\cevns rate in our previous WIMP search region~\cite{xenonnt_wimp} is 0.2 events/(t$\times$yr). To increase the rate of detected \beight\,\cevns in this search, the signal acceptance is improved by lowering two thresholds. First, the S2 signal threshold is reduced from 200\,PE in the WIMP search to 120\,PE in this search. Second, the S1 coincidence requirement was lowered from threefold coincidence to twofold coincidence, now minimally requiring only two PMTs with hits within $\pm \SI{50}{\nano\second}$ around the maximal amplitude of the S1 waveform. The reduced thresholds lead to an expected \beight\,\cevns detection rate of 3.7(3.3)\,events/(t$\times$yr) in SR0(1), a factor of $\sim$17 larger than in the WIMP search.

The ROI in this analysis is defined to be two or three hits for \Sone signals and (120, 500)\,PE for \Stwo signals. The upper bound of the \Stwo area range is set to retain most of \cevns signal and to remove electronic recoil \,(ER) backgrounds from $\upbeta$ and $\upgamma$ radiation, which have higher ratios of S2 to S1 than NRs~\cite{Aprile:2006kx}. S1 signals with more than three hits are rarely produced by \beight~\cevns and such events are therefore not included in this analysis. Events in the ROI are blinded except those with radial positions larger than 63.0\,cm, which are used to model the surface events produced by $^{210}$Pb plate out on the TPC wall~\cite{XENON:2018voc_1t_1ty}, and are not part of the dataset for the search. Threefold events were unblinded in the SR0 WIMP search~\cite{XENONnT:2023dvq}, which contributes to $\leq3$\% of total \beight~\cevns rate since twofold events dominate and SR1 has more exposure.

Cuts based on the features of S1 and S2 peaks, inherited from~\cite{XENON:2024qgt}, are employed to ensure the quality of the reconstructed events. \Sone signals composed of at least two hits are required to be larger than 1\,PE. \Sone signals up to 4\,PE are accepted in size per PMT. S2 signals must be detected by both PMT arrays with a reasonable signal fraction of around 75\% in the top array. S2 signals detected on the top array are also required to follow the expected pattern from the optical response of XENONnT. Events with multiple S2 signals are rejected to suppress the neutron background. As in~\cite{XENON:2024qgt}, events found in coincidence with either MV or NV are rejected.

\itsec{Backgrounds}
This analysis considers accidental coincidence\,(AC), surface, neutron, and ER background components, as in the search for solar \beight~\cevns signals with the XENON1T detector~\cite{xenon1t_b8,XENON:2024xgd}. 
The AC is the dominant background, formed by accidentally paired ``isolated" S1 and S2 signals. The accidental pileup rate of these isolated S1 and S2 signals within the maximum drift time is significant, reaching several hundred events per day before mitigation measures are applied.

The primary source of the isolated S1 and S2 signals in the \beight~\cevns search ROI are delayed signals after high-energy~(HE) interactions. These interactions, with characteristic S2 areas larger than \SI{10000}{PE} induced predominantly by $\upgamma$ rays from the materials' radioactivity, are known to contaminate their subsequent time interval with single photoelectron PMT hits and small S2 signals.
This phenomenon has been observed in many LXe detectors~\cite{xenon1t_b8, LZ:2022lsv-wimp, PandaX:2022aac-b8}. 
While the physical mechanism is still under investigation~\cite{Kopec:2021ccm-few-e, Sorensen:2017ymt-few-e}, the AC background can be modeled by data-driven simulation, after applying dedicated cuts to remove the isolated peaks correlated with their preceding HE peaks.  

The impact on an isolated signal by a preceding HE event is quantified by the ratio of $\Stwo_\mathrm{pre}$ to $\Delta t_\mathrm{pre}$, where $\Stwo_\mathrm{pre}$ is the S2 area of the HE event and $\Delta t_\mathrm{pre}$ is the time between the HE event and the isolated signal. All the HE events 1 sec before the isolated signal are considered and the event with the largest ratio of $\Stwo_\mathrm{pre}$ to $\Delta t_\mathrm{pre}$ (defined as \prevstwodt) is identified as the most influential one on the isolated signal rate. Cuts are then applied on \prevstwodt to minimize the isolated signal rate. A time window of \SI{2.2}{\milli\second}~(one maximum drift time) is vetoed after any HE interaction in SR0. In SR1, due to the increased photoionization rate, the veto window is extended to \SI{4.4}{\milli\second}. The cut on \prevstwodt for 2~(3)-hit S1 signals is less than $10.1$~($38.2$)~PE/\textmu s, effectively reducing isolated S1 rates by more than 80\%~(50\%) while accepting 87\%~(96\%) of \beight~\cevns signals. 
Localized bursts of intense single-electron\,(SE) emission observed in SR0~\cite{xenonnt_wimp} appear more frequently in SR1, contributing also to the isolated \Stwo signals. 
For isolated S2 signals, correlations with preceding HE events and the localized SE burst in ($X, Y$) position are utilized, accounting for the uncertainty in position reconstruction. 
Two-dimensional cuts in time and position are developed, effectively rejecting over 50\% of isolated S2 signals while accepting around 96\% of \beight~\cevns signals.

After all the cuts, the average isolated S1 and S2 signal rates in SR0~(SR1) are 2.3~(2.2) Hz and 18~(26)\,mHz, respectively.
By injecting simulated \beight~\cevns signals at random times and positions into the real data, the overall acceptance of these cuts is evaluated to be 75\%~(85\%) for 2-~(3)-hit signals. The isolated S1 and S2 waveforms are then sampled and assigned a random drift time before being merged into artificial AC events. Facilitated by~\cite{axidence}, the simulation improved compared to \cite{xenon1t_b8} in preserving the \prevstwodt spectrum and modeling the time dependence to minimize the systematic uncertainties of the AC model.

Two boosted decision tree\,(BDT) classifiers are developed to distinguish between \beight~\cevns signals and the AC background events. The output scores from these classifiers are used as analysis dimensions in the final likelihood.
The distributions of S1 photons of \beight~\cevns signals in time and across the PMT arrays differ from those of the isolated S1 signals induced by a random pileup of PMT hits. Features from these distributions are therefore combined in an \Sone BDT score. Another BDT assesses the \Stwo signal shape and the time between the S1 and S2 signals, which in \beight~\cevns signals are correlated due to diffusion of the drifting electron cloud, but this correlation is absent for the AC background. A cut on the \Stwo BDT score is applied to reject about 90\% of the AC background events while retaining more than 80\% of the signal events.

The S2 pulse shape changes close to the perpendicular supporting wires~\cite{XENONnT:2023dvq,xenonnt_wimp}, so applying the S2 BDT cut to those events would introduce systematic errors in signal acceptance. 
Consequently, events close to the perpendicular wires are excluded from the analysis. Because of the S2-area-dependent position resolution, this leads to an S2-area-dependent reduction in the S2 acceptance rather than a reduction of the fiducial mass. 
Simulated S1 and S2 waveforms~\cite{fuse} are used to assess the acceptance loss due to cuts. The difference between acceptances estimated by simulated events and calibration data is smaller than 10\%, which is assigned as the uncertainty on the total acceptance. \fref{fig:detection_acceptance} shows the total acceptance for \Sone- and \Stwo-based cuts as function of NR energy.

AC-rich datasets are selected to validate the AC background model, including events with unphysically long drift times, calibration datasets featuring a high rate of isolated peaks, and an AC sideband mainly made of events rejected by the S2 BDT cut. 
These validations are performed with a binned likelihood goodness of fit\,(GOF) test in all the same dimensions as used in the statistical inference to search for \beight~\cevns signals, including c\Stwo, \prevstwodt, \Sone BDT score, and \Stwo BDT score. 
In all these validation datasets, good agreements between AC prediction and observation in the \beight\,\cevns ROI is achieved, constraining systematic uncertainties on the AC rate to be below 5\%. Conservatively, the systematic uncertainty of the AC background is solely estimated from the AC sideband, which is unblinded only after the AC prediction and event selections are both fixed. The AC model passed the binned likelihood GOF test with the sideband data at a p-value of 0.16. 
The AC background uncertainty for SR0 (SR1) is 9.0\%~(5.8\%), based on statistical uncertainties from the AC sideband data. The expected numbers of AC background in SR0 and SR1 are (7.5$\pm$0.7) and (17.8$\pm$1.0), respectively. Details about the AC sideband unblinding are provided in \aref{app:appendixA}.

Surface events produced mainly by $^{210}$Pb plate out on the TPC wall have reduced S2 signals~\cite{XENON:2018voc_1t_1ty}, which could lead to leakage of events into the ROI. A data-driven approach is adopted to derive the radial distribution of this background. Because of the limited statistical data, deriving and validating the data-driven model across all four analysis dimensions is currently unfeasible. Consequently, the outer radius of the FV for SR0 (SR1) is set at \SI{60.15}{\centi\meter} (\SI{59.60}{\centi\meter}), such that surface events are expected to be less than 0.12~(0.23), respectively. At this level, this background can be safely neglected without risk of signal-like mis-modeling in the \beight~\cevns search according to a dedicated toy Monte Carlo~(MC) study.

Radiogenic neutrons originating from the detector materials are modeled using the framework of ~\cite{XENON:2024xgd} with neutron spectra from updated knowledge of the detector material radioactivity. The prediction for SR0 and SR1 are ($0.13 \pm 0.07$) and ($0.33 \pm 0.19$) events, respectively. The rate uncertainty of 58\% is derived from neutron candidates in SR0 tagged by the NV. In the CEvNS ROI, the NV and MV tagged one event each after a dedicated unblinding, which is in agreement with the expected number of events vetoed by accidental coincidence between the TPC and the veto detectors.

The ER background is composed mainly of $\beta$ decays from radioactive impurities such as $^{214}$Pb and $^{85}$Kr, and electrons scattered by external $\upgamma$ rays and solar neutrinos\,\cite{XENON:2024xgd}. The shape of the ER background in the \beight~\cevns ROI is generated by \cite{appletree} with emission model fit to the $^{220}$Rn calibration data~\cite{XENON:2024xgd}. The rate of ER background events is derived by fitting the events with ER energy above 20\,keV, assuming a flat ER spectrum. However, the emission model in low energy has large systematical uncertainty. If using the light and charge yields from the Noble Element Simulation Technique(NEST)~\cite{Szydagis:2022ikv,nest}, the expected ER rate is 10 times lower. To account for this discrepancy, a 100\% uncertainty is assigned to the ER rate. Consequently, the assumed ER background in SR0 and SR1 is taken to be at most $0.13 \pm 0.13$ and $0.56\pm0.56$ events, respectively. Measurement of the light and charge response in XENONnT with a 0.27\,keV calibration using a \arsq electron capture~(EC) source, which will be introduced in a future publication, also confirms the nominal rate of the ER background is a conservative choice. 

\begingroup
\renewcommand{\arraystretch}{1.5}
\begin{table}[ht]
    \centering
    \caption{The expected and best-fit number of events from signal and background components in the ROI. The uncertainty in the expectation accounts for contributions from signal detection efficiency, \ly, and \qy. The uncertainties of background expectations correspond to the width of the Gaussian constraints in the fit, the \beight signal is not constrained. }
    \vspace{0.1cm}
    \begin{tabular}{
        >{\centering}m{2.5cm} 
        >{\raggedleft}m{1.0cm}
        c
        >{\raggedright}m{1.0cm}
        >{\raggedleft}m{1.0cm}
        c
        >{\raggedright\arraybackslash}m{1.0cm}
    }
    \hline\hline
    Component & \multicolumn{3}{c}{\makecell{Expectation}} & \multicolumn{3}{c}{\makecell{Best-fit}} \\
    \hline
    AC (SR0)  & 7.5&$\pm$&0.7& 7.4&$\pm$&0.7 \\
    AC (SR1)  & 17.8&$\pm$&1.0& 17.9&$\pm$&1.0 \\
    ER        & 0.7&$\pm$&0.7& \multicolumn{3}{c}{$0.5^{+0.7}_{-0.6}$} \\
    Neutron   & \multicolumn{3}{c}{$0.5^{+0.2}_{-0.3}$} & 0.5&$\pm$&0.3 \\
    \hline
    Total background & \multicolumn{3}{c}{$26.4^{+1.4}_{-1.3}$}& 26.3&$\pm$&1.4 \\
    \beight   & \multicolumn{3}{c}{$11.9^{+4.5}_{-4.2}$} & \multicolumn{3}{c}{$10.7^{+3.7}_{-4.2}$} \\
    \hline
    Observed & \multicolumn{6}{c}{37} \\
    \hline\hline
    \end{tabular}
    \label{tab:backgrounds}
\end{table}
\endgroup

\itsec{Statistical Inference}
\prevstwodt, \Sone BDT score, \Stwo BDT score, and cS2 are the four dimensions used to discriminate between the \beight~\cevns signal and the dominating AC background. The background and signal models are coarsely binned, with three bins in each of the four analysis dimensions for a total of 81 bins. A four-dimensional binned likelihood analysis is performed. The bins are chosen to have the same expected number of AC background events in the projection of each dimension. The chance for mis-modeling of the AC background due to the limited number of isolated S1 and S2 peaks is negligible, as validated via toy MC simulations.

The extended likelihood function is constructed as
\begin{equation}
    \centering
    \lagr(\mu,\nuiss) = \prod_{i=0,1} \lagr_{i}(\mu,\nuiss)\times \prod_m \lagr_m(\nuis_m),
    \label{eqn:ll_total}
\end{equation}

\noindent where the parameter of interest $\mu$ can either be the solar \beight neutrino flux\,($\Phi$), or the flux-weighted \cevns cross section on Xe\,($\sigma_\cevns$). $\nuiss$ are the nuisance parameters, $i$ iterates through the two science runs, and $m$ iterates through the nuisance terms: the constraints on $\tly$ and $\tqy$, the signal acceptance uncertainty and the uncertainties in the rates of the AC, neutron, and ER backgrounds. The nuisance parameters $\nuis_m$ are constrained via external measurements, modeled by Gaussian pull terms $\lagr_m(\nuis_m)$. The models of \beight~\cevns and neutrons change in shape and expectation value with $\tly$ and $\tqy$. The AC background rates are independent between science runs, while all other parameters are coupled.

The \beight~\cevns discovery significance and the construction of a confidence interval for the \beight neutrino flux are computed using a test statistic\,$q_{\mu}$ based on the profile log-likelihood ratio as in~\cite{Wilks:1938dza,XENON:2024xgd}. The critical region for the confidence interval construction and expected discovery significance are computed with toy MC simulations using \cite{alea}. Consistency between the model and data is evaluated by a combination of four binned likelihood GOF tests performed on the four one-dimensional projections, combining SR0 and SR1. The p-values are computed based on the distribution of the binned likelihood GOF test statistic obtained via toy MC simulations. A threshold of 0.013 is selected for each test to obtain a 95\% confidence limit\,(CL) for the final combined test. The test is defined before unblinding and its suitability to reject mis-modeling is assessed using toy MC simulations.

The strategy to report the result from the \beight~\cevns search is decided before unblinding. A Feldman-Cousins construction~\cite{Feldman:1997qc} is used to constrain the solar \beight neutrino flux and the \cevns cross section $\sigma_\cevns$ without setting a threshold on p-values for reporting a two-sided measurement. 
The expected \beight~\cevns signal under the nominal emission model is $11.9^{+4.5}_{-4.2}$ events, with the uncertainty originating from S1 and S2 acceptances and detector response to low-energy NRs. The expected background is $26.4^{+1.4}_{-1.3}$ events, dominated by the AC background, as shown in Tab.~\ref{tab:backgrounds}. With the final background prediction summarized in~\tref{tab:backgrounds}, the probability of obtaining a $\geq2\sigma$~(3$\sigma$) discovery significance with this dataset is estimated to be 80\%~(48\%) using toy MC simulations.

\begin{figure}[t]
    \centering
    \includegraphics[width=0.95\columnwidth,left]{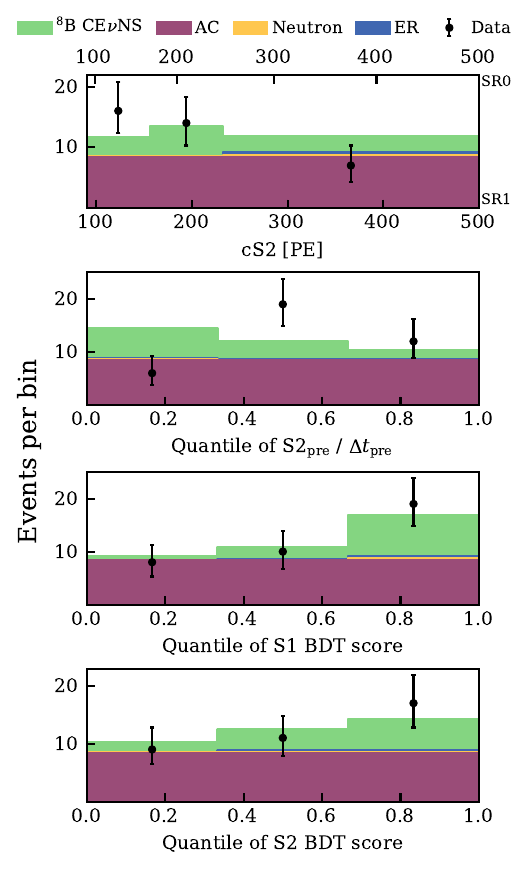}
    \caption{Distributions of best-fit signal and background, together with the data in the projected analysis dimensions, summing both science runs. The observed number of events with Poisson uncertainties in each bin is shown in black. The \beight~\cevns signal is represented by the light green histogram on top of the backgrounds, which are indicated by purple\,(AC), blue\,(ER), and yellow\,(neutron) histograms. As the bin edges on each analysis dimension vary from SR0 to SR1, the plot for c\Stwo is shown in double axis, and the other dimensions are shown in quantiles of the AC background for the summed results.
}
    \label{fig:best_fit}
\end{figure}

Before unblinding the \beight~\cevns search data, the signal and background modeling are validated in the four-dimensional space by measuring the \ly of the \arsq L-shell electron capture ER signal at \SI{0.27}{\kilo\electronvolt}, where \qy is constrained~\cite{XENON:2022ivg} but \ly has not yet been measured. The background in the \arsq data at this low-energy region is dominated by the AC background due to the high rate of isolated S2 signals. The \ly of \arsq L-shell is measured by fitting the \arsq calibration data~\cite{XENON:2022ltv} with the \arsq signal and the AC background. This fitting is analogous to the search for \beight~\cevns signals in terms of the signal dependence on the light and charge yields, the dominant background, and the energy region. Using approaches on the signal and background modeling comparable to the \beight~\cevns search, the best-fit of the \arsq signal model and the AC background is consistent with the data in all of the four analysis dimensions. More information about this validation is described in \aref{app:appendixB}.

\itsec{Results} After unblinding, 9 and 28 events are observed in SR0 and SR1, respectively. The observed number of events is consistent with the expected \beight~\cevns signal on top of the background. The best-fit values of background components and \beight~\cevns signal from the unconstrained fit are also shown in \tref{tab:backgrounds}. The best-fit nuisance parameters $\nuiss$ are all within $\pm\,0.3\,\sigma$ constrained by the external measurements. The background-only hypothesis, with no \beight~\cevns signals, is disfavored with a p-value of 0.003, corresponding to a statistical significance of 2.73\,$\sigma$. 

The distributions of the observed 37 events and the best-fit model projected to each analysis dimension are shown in \fref{fig:best_fit}. A detailed plot showing the SR0 and SR1 results separately is presented in \aref{app:appendixC}. The p-values in c\Stwo, \Sone BDT score, and \Stwo BDT score show a good match between the unconstrained best-fit model and observations. The p-value in the \prevstwodt is 0.008, indicating a potential mis-modeling. 
No other indication of possible mis-modeling is found by inspecting the individual events in the dataset or the AC sideband data.
Abandoning \prevstwodt in the statistical inference would lead to a larger best-fit \beight~\cevns signal of 13.1 events with a statistical significance of $3.22\,\sigma$. In addition, two tests of overdensity in ($X, Y$) space were defined before unblinding, although not part of the analysis dimensions. One returned a p-value below the threshold of 0.018, prompting checks including inspection of event distributions in all cut spaces that show no indication of mis-modeling.

\begin{figure}[t]
    \centering
    \includegraphics[width=0.95\columnwidth,left]{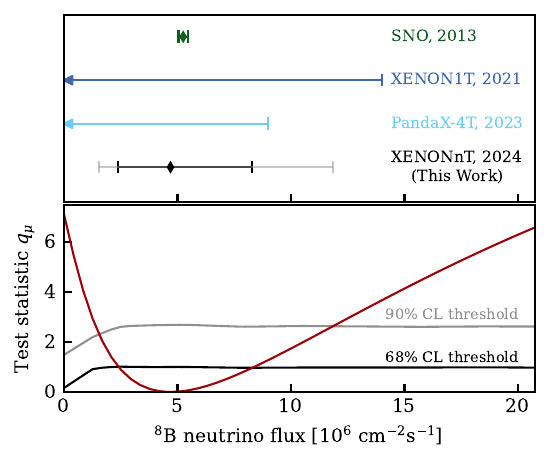}
    \caption{
    Constraints on solar \beight neutrino flux. Top: the 68\%~(90\%)\,measurement of solar \beight neutrino flux from this work is shown in black\,(gray). The 68\%\,CL measurement from SNO~\cite{SNO:2011hxd}, and 90\% CL upper limits from XENON1T~\cite{xenon1t_b8} and PandaX-4T~\cite{PandaX:2022aac-b8} are also shown. Bottom: the solid red line shows the profile likelihood ratio test statistics $q_{\mu}$ as a function of solar \beight neutrino flux. The constraints are derived with Feldman-Cousins construction at 68\%~(90\%)\,CL, indicated by the black\,(gray) curve. }
    \label{fig:b8_flux}
\end{figure}

\begin{figure}[t]
    \centering
    \includegraphics[width=1.0\columnwidth,left]{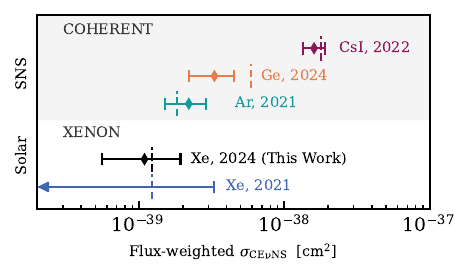}
    \caption{Measurements of the flux-weighted \cevns cross section $\sigma_\cevns$. The measurement using Xe nuclei solar \beight neutrinos from this work is shown in black. The 90\% CL upper limit from XENON1T~\cite{xenon1t_b8} is shown in blue. The measurements with neutrinos from the SNS by the COHERENT collaboration using CsI~\cite{COHERENT:2021xmm}~(red), Ar~\cite{COHERENT:2020iec}~(green) and Ge~\cite{Adamski:2024yqt}~(orange) nuclei are also shown. For comparison, the SM predictions are shown by vertical dashed lines. 
    }
    \label{fig:b8_ross_section}
\end{figure}

Assuming the flux-weighted \cevns cross section $\sigma_\cevns$ predicted by the SM, \fref{fig:b8_flux} shows the XENONnT constraint on the solar \beight neutrino flux of \beightflux at 68\% CL. 
With the solar \beight neutrino flux being constrained by SNO~\cite{SNO:2011hxd}, 
\fref{fig:b8_ross_section} shows the first measurement of the flux-weighted \cevns cross section $\sigma_\cevns$ on Xe as $(1.1^{+0.8}_{-0.5})\times10^{-39}\,\mathrm{cm}^2$, consistent with the SM prediction of 1.2$\times10^{-39}\mathrm{cm}^2$. Since the momentum transferred from a solar \beight neutrino to a Xe nucleus is $\leq$ \SI{20}{\mev}/c, this measurement is less sensitive to uncertainties in the nuclear form factor compared to \cevns measurements made by the COHERENT collaboration with neutrinos produced by the SNS~\cite{COHERENT:2021xmm}. The measurements of the flux-weighted \cevns cross section on CsI~\cite{COHERENT:2021xmm}, Ar~\cite{COHERENT:2020iec}, and Ge~\cite{Adamski:2024yqt} nuclei by the COHERENT Collaboration are shown in \fref{fig:b8_ross_section} for comparison. Because of the lower average energy, the solar \beight neutrino flux-weighted \cevns cross section is the lowest one measured to date.

\itsec{Summary} We performed a blind search for NR signals from solar \beight neutrinos via \cevns with XENONnT using data from two science runs with a combined exposure of 3.51\,t$\times$yr. By lowering the S1 and S2 thresholds, we are able to include NR signals as low as 0.5\,keV. Various techniques are developed to reduce the dominant AC background. Various calibrations, including \ybe and \arsq, are performed to understand the detector response, signal, and background modeling. The data disfavor the background-only hypothesis at 2.73\,$\sigma$. The unconstrained best-fit number of \beight~\cevns signals is $10.7^{+3.7}_{-4.2}$, consistent with the expectation of $11.9^{+4.5}_{-4.2}$ events, based on the measured solar \beight neutrino flux from SNO~\cite{SNO:2011hxd}, the theoretical \cevns cross section with Xe nuclei~\cite{Barranco:2005yy}, and the calibrated detector response to low-energy NRs in XENONnT. Thus, the measured solar \beight neutrino flux is \beightflux, consistent with SNO, and the measured neutrino flux-weighted \cevns cross section on Xe is $(1.1^{+0.8}_{-0.5})\times10^{-39}\,\mathrm{cm}^2$, consistent with the SM prediction. As XENONnT continues to take data, more precise measurements are expected in the future.

\itsec{Acknowledgements}
We would like to thank the COHERENT Collaboration for providing data points and predictions for the measurement of flux-weighted \cevns cross section $\sigma_\cevns$ at the SNS.

We gratefully acknowledge support from the National Science Foundation, Swiss National Science Foundation, German Ministry for Education and Research, Max Planck Gesellschaft, Deutsche Forschungsgemeinschaft, Helmholtz Association, Dutch Research Council (NWO), Fundacao para a Ciencia e Tecnologia, Weizmann Institute of Science, Binational Science Foundation, Région des Pays de la Loire, Knut and Alice Wallenberg Foundation, Kavli Foundation, JSPS Kakenhi and JST FOREST Program ERAN in Japan, Tsinghua University Initiative Scientific Research Program, DIM-ACAV+ Région Ile-de-France, and Istituto Nazionale di Fisica Nucleare. This project has received funding/support from the European Union’s Horizon 2020 research and innovation program under the Marie Skłodowska-Curie grant agreement No 860881-HIDDeN.

We gratefully acknowledge support for providing computing and data-processing resources of the Open Science Pool and the European Grid Initiative, at the following computing centers: the CNRS/IN2P3 (Lyon - France), the Dutch national e-infrastructure with the support of SURF Cooperative, the Nikhef Data-Processing Facility (Amsterdam - Netherlands), the INFN-CNAF (Bologna - Italy), the San Diego Supercomputer Center (San Diego - USA) and the Enrico Fermi Institute (Chicago - USA). We acknowledge the support of the Research Computing Center (RCC) at The University of Chicago for providing computing resources for data analysis.

We thank INFN Laboratori Nazionali del Gran Sasso for hosting and supporting the XENON project.

\vspace{0.5cm}

\itsec{Note Added}
Recently, we noticed the results of the \beight neutrino flux measurement from the PandaX Collaboration with a similar statistical significance in~\cite{PandaX:2024muv}.

\vspace{0.5cm}

\begin{figure}[ht]
    \centering
    \includegraphics[width=0.95\columnwidth,left]{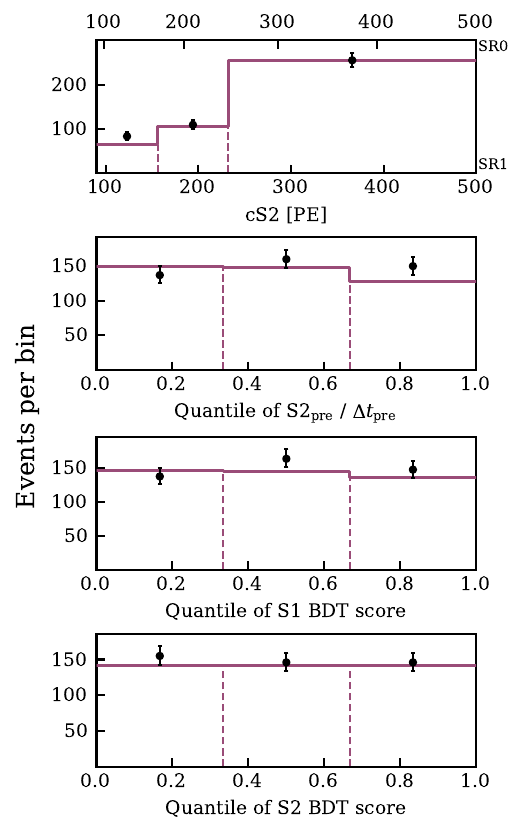}
    \caption{Distributions of expected AC background in AC sideband and the observed data in the projected analysis dimensions. Both expectation and observation have \Stwo larger than \SI{120}{\PE}. The expected AC background is shown in the purple histogram. The observed number of events with Poisson uncertainties in each bin is shown in black.
    }
    \label{fig:sideband}
\end{figure}

\appendixsection{AC Sideband Validation} The AC sideband validation is also performed with a blind analysis, before unblinding the \beight~\cevns search data. After the AC event selection and prediction are both fixed, the SR0 and SR1 AC sideband datasets are unblinded. With the initial \Stwo threshold of \SI{100}{PE}, 133~(416) events are observed in SR0~(SR1) with an expectation of 135.9~(368.2). With the four-dimensional binned likelihood GOF test, the prediction and the observation in SR0 show an acceptable agreement. However, the test on SR1 showed a mismatch with a p-value of 0.03. All the analysis dimensions are inspected and the mismatch is only present below \SI{120}{\PE} in \Stwo, suggesting that the mismatch in SR1 is most likely due to the increase in photoionization. 
The S2 thresholds for both SR0 and SR1 are thus conservatively increased for the \beight~\cevns search data, with minor loss in the discovery potential of the solar \beight~\cevns signals. The final prediction of the AC background and observations in the AC sideband are shown in \tref{tab:acsideband}. The projection of the four analysis dimensions with the same binning used in the \beight~\cevns search in sideband data with the \Stwo larger than \SI{120}{\PE} in both SR0 and SR1 data are shown in \fref{fig:sideband}.

\begin{table}[ht]
    \centering
    \caption{AC sideband validation. The expected and observed numbers of events are for a 120\,(100) PE \Stwo threshold. }
    \begin{tabular}{
            >{\centering}m{2cm}
            >{\centering}m{2cm}
            >{\centering}m{2cm}
            >{\raggedright\arraybackslash}m{2cm}
        }
        \hline\hline
        Science run  & Expectation & Observation & p-value \\
        \hline
        SR0 & 122.7\,(135.9) & 121\,(133) & 0.33\,(0.74) \\
        SR1 & 302.5\,(368.2) & 326\,(416) & 0.16\,(0.03) \\
        \hline\hline
    \end{tabular}
    \label{tab:acsideband}
\end{table}

\begin{figure}[ht]
    \centering
    \includegraphics[width=0.95\columnwidth,left]{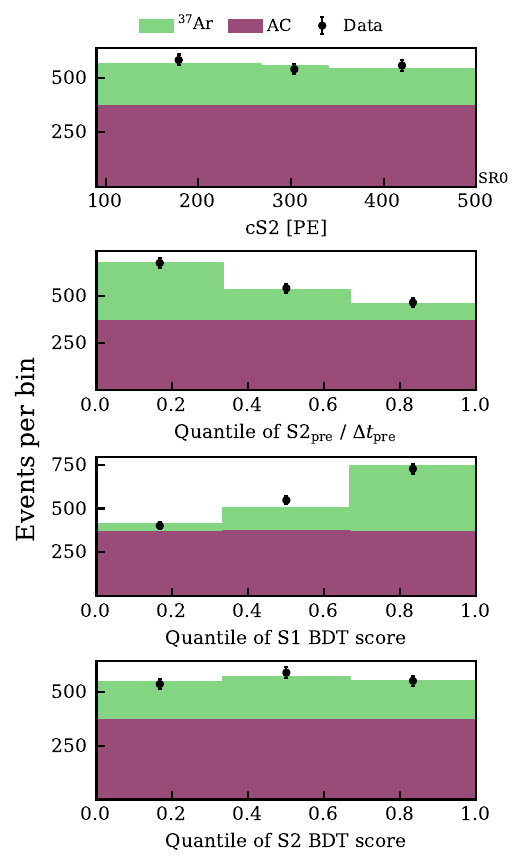}
    \caption{Distributions of the best-fit AC background, \arsq L-shell EC signal, and the observed data in the \beight~\cevns analysis dimensions. The observed number of events with Poisson uncertainties in each bin is shown in black. The \arsq signal\,(AC background) is shown by the green\,(purple) histogram. 
    }
    \label{fig:arsq}
\end{figure}

\appendixsection{Modeling validation} The signal and background modeling is validated by the measurement of \ly of \arsq L-shell EC, which is performed with a blind analysis. The AC background in this measurement is estimated to be 1062$\pm$53 based on a similar modeling approach to that in the \beight~\cevns search. After unblinding, 1676 events are observed. The observed events above the expected AC background are strongly validated by a four-dimensional GOF test, yielding a p-value of 0.92. \fref{fig:arsq} shows the observed events in the same analysis dimensions as the \beight~\cevns search along with the AC background and the best-fit \arsq L-shell EC signal during the \arsq calibration. The measurement will be presented in a future publication.

\begin{figure*}
  \centering
  \includegraphics[width=\textwidth]{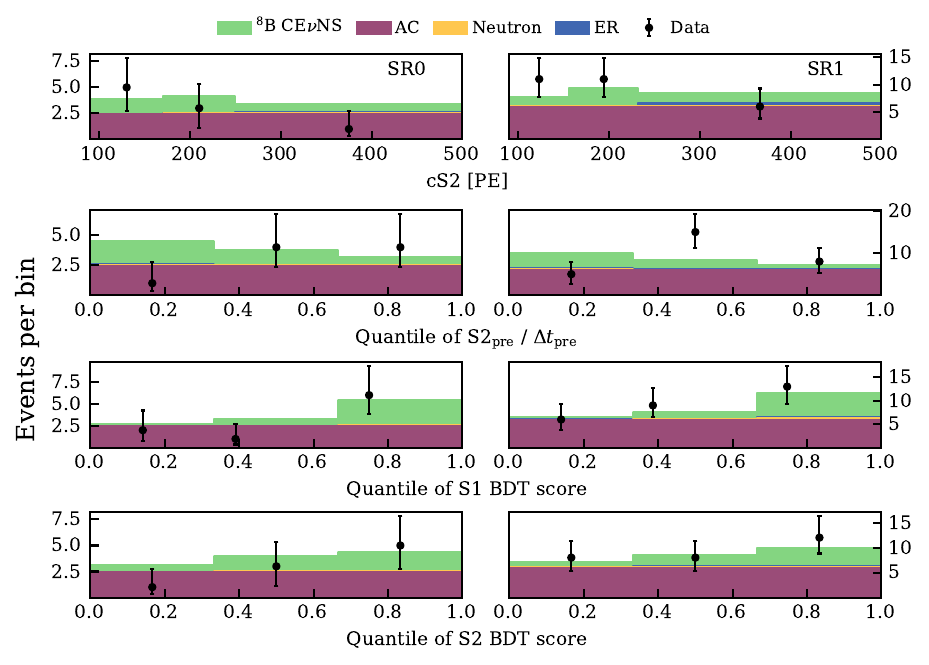}
  \caption{Distributions of best-fit signal and background, together with the data in the projected analysis dimensions, with SR0 and SR1 shown in the left and right column, respectively. The observed number of events with Poisson uncertainties in each bin is shown in black. The \beight~\cevns signal is represented by the light green histogram on top of the backgrounds, which are indicated by purple\,(AC), blue\,(ER), and yellow\,(neutron) histograms. }
  \label{fig:result_breakdown}
\end{figure*}

\appendixsection{Separate SR0/SR1 best-fit results} 
The distributions of the observed events in SR0 and SR1 and the corresponding best-fit model projected to each analysis dimension are shown individually in Fig.~\ref{fig:result_breakdown}.

\bibliography{bibliography}

\end{document}